# Bridging the Bubbles: Connecting Academia and Industry in Cybersecurity Research

**Rasha Kashef***
**Electrical, Computer, and Biomedical Engineering, Toronto Metropolitan University**
**Rogers Cybersecure Catalyst Fellow**

**Monika Freunek***
**Lighthouse Science Consulting and Technologies Inc.**
**Rogers Cybersecure Catalyst Fellow**

**Jeff Schwartzentruber**
**eSentire Inc.**
**Rogers Cybersecure Catalyst Fellow**

**Reza Samavi**
**Electrical, Computer, and Biomedical Engineering, Toronto Metropolitan University**
**Vector Institute of Artificial Intelligence**
**Rogers Cybersecure Catalyst Fellow**

**Burcu Bulgurcu**
**Ted Rogers School of Management, Toronto Metropolitan University**
**Rogers Cybersecure Catalyst Fellow**

**AJ Khan**
**Vehiqilla Inc.**
**Rogers Cyber Secure Catalyst Fellow**

**Marcus Santos**
**Computer Science, Toronto Metropolitan University**
**Rogers Cybersecure Catalyst Academic Director**

**\*: Corresponding Authors.**

## Abstract

There is a perceived disconnect between how ad hoc industry solutions and academic research solutions in cyber security are developed and applied. Why is that? Is there a difference in philosophy in how solutions to cyber security problems are developed by industry and by academia? What could academia and industry do to bridge this gap and speed up the development and use of effective cybersecurity solutions? This paper provides an overview of the most critical gaps and solutions identified by an interdisciplinary expert exchange on the topic. The discussion was held in the form of the webinar "Bridging the Bubbles: Connecting Academia and Industry in Cybersecurity Research" in November 2022 as part of the Rogers Cybersecure Catalyst webinar series. Panelists included researchers from academia and industry as well as experts from industry and business development. The key findings and recommendations of this exchange are supported by the relevant scientific literature on the topic within this paper. Different approaches and time frames in development and lifecycle management, challenges in knowledge transfer and communication as well as heterogeneous metrics for success in projects are examples of the evaluated subject areas.

**Keywords**: Cyber Security R&D, Industry Solutions, Academic Research, Development, Knowledge Transfer, Commercialization

## 1. Introduction

The global cyber security workforce has been estimated to have reached a record 4.7 million people in 2022 [1]. However, the estimated gap in the additional required workforce amounts to another 3.4 million. So far, education, training and research in cyber security are emerging topics on a global scale. At the time of writing, less than 20 Canadian universities offer master programs in cyber security topics even when using a broad definition of cyber security [2]. On the other hand, constantly evolving new threats and the asymmetric situation between attackers and defenders in cyberspace require a rapid and constant research environment, where not only current threats and mitigation strategies are identified and developed, but also future threats and their mitigation are analyzed and evaluated. Compared to other scientific fields, this yields a unique, challenging research environment where both a shortage of qualified researchers and a need for rapid development and field implementation collide. Cyber security research is further unique in that it requires both a fundamental understanding of advanced disciplines within computer science, mathematics, statistics, and engineering, such as machine learning, encryption, network theory and hard- and software knowledge, but also timely access and practical understanding of real-world applications, systems, and operational workflows within real organizations. Some of these application-oriented aspects require an interdisciplinary approach. For these reasons, it has been suggested that cybersecurity studies and education should be addressed and designed as a "meta discipline" and based on cross-sectoral partnerships ([3]-[5]). However, cyber systems and data are protected from outside access in many real-world applications, including university research activities. This poses a barrier to researchers in identifying and understanding current cyber challenges in real application environments, but also to industry and the public sector, as solutions from research are not readily accessible or directly suitable to the application environment. Consequently, security research for industrial applications is often performed within organizations and mostly focuses on selected specific areas. Research activities are provided by external private companies or are not pursued at all.

Cybersecurity research in academia and industry faces several structural challenges that lead to a disconnect between the sectors. To improve the understanding of this challenge and identify potential solutions, the research fellows of the Rogers Cybersecure Catalyst of Toronto Metropolitan University held a webinar with panelist discussions on the disconnect between cybersecurity research in academia and industry [6]. The panelists: the Director of Innovation and Policy of the Rogers Cybersecure Catalyst of Toronto Metropolitan University, Sumit Bhatia, the Senior Vice President of the Cyber Security Product Engineering of Blackberry, Shiladitya Sircar, Dr. Reza Samavi, associate professor in the Electrical, Computer, and Biomedical Engineering, Toronto Metropolitan University and a faculty affiliate with the Vector Institute of Artificial Intelligence, and Dr. Jeff Schwartzentruber, a senior Machine Learning Scientist at eSentire, Adjunct at Dalhousie University in the Faculty of Computer Science and Special Faculty at the University of Guelph in the School of Computer Science, were invited and contributed to the discussion with their background and experience in cyber security innovation for industrial and academic settings. This work presents the results of this event with a focus on solutions. In addition, supporting findings from the relevant scientific literature are provided. The paper focuses on four selected major questions. First, the influence of the different development approaches in research and industry is discussed, where an evolutionary process is identified to be the most promising for practical applications. While academia can provide expertise, the most important and novel research questions in cybersecurity can be identified the most reliably in the field of practice itself. Different solutions range from specific and often secretive ad hoc solutions in the industry to scientifically peer-reviewed and publicly available research outcomes. The second question evaluates the influence of the requirements on lifecycles and development cycles, especially regarding the time factor until the implementation of a new solution, and on opportunities to align the methodologies between industry and academia. Examples from the automotive sector are outlined. The third thematic section identifies solutions to improve knowledge transfer in the cybersecurity field and approaches to overcoming so-called knowledge silos with respect to the unique working conditions of the cybersecurity field. An important component of this challenge is communication, which is the main topic evaluated in detail in the last guiding question. Development goals, requirements and metrics for success are discussed with a focus on the commercial use and implementation of cybersecurity solutions. The paper concludes with a summary of the key points of the discussion, the significant challenges identified during the webinar and its key recommendations on how to empower the current and future cybersecurity research in industry and academia.

## 2. Webinar Discussion Questions

### 2.1 A disconnect between ad hoc industry solutions and research: how could solutions out of research be sped up in their transition to the industry?

A significant driver of public policy with respect to University-Industry Collaborations (UICs) is to generate innovative IP that results in increased economic growth [7], [8]. The positive economic impacts of UICs have been well studied [9], with many public policies supporting the creation of networks, clusters, and institutions [10]. Research has shown that the economic potential of UICs is maximized through the collaboration of industry, universities, and government agencies [11]. However, realizing the economic benefits of UICs depends on their ability to generate a fruitful innovation and the speed at which it can be commercialized. One well-documented barrier to UICs is the velocity at which academic research can be commercialized [12]. The industry partners' short-term directives do not align with academic research's longer and more exploratory nature [11]. The friction caused between these two temporal scales reduces the efficacy of UICs. The interplay of the various barriers and enablers of UICs results in a complex and confounding discussion concerning commercialization. For the sake of brevity, the following section will only focus on discussing factors affecting the 'speed' at which industry research can be transitioned to industry.

Within UIC literature, this translation of research from the academic partner to the industry (and vice-versa) is known as Knowledge Transfer [13]. However, it should be noted that knowledge transfer does not equal commercialization. For example, it is unreasonable to expect that the academic partner has the insight and capability to deploy and maintain the generated innovation within the commercial environment of the industry partner. Although knowledge transfer overlaps during a UIC project, they can also be viewed as independent in many contexts. One significant theme affecting UICs is the transparency between the three parties – the university, industry partner and public sector [11], [12]. One area that is especially convoluted during a UIC is understanding the motivations of the various stakeholders and the critical success factors [11]. From a university perspective, the primary motivation is fundraising for project financing, linked to several university-wide and researcher-specific Key Performance Indicators (KPIs) [12]. Comparatively, the motivation of the industry partner is to have access to highly qualified personnel and KPIs typically linked to revenue generation or cost reduction [12]. Although each respective stakeholder understands these motivations well, they are typically not well received between parties. This difference is especially polarizing for industry stakeholders with little understanding of the rigor required by academic research and academics with little industry experience. Thus, the first step in accelerating the rate of knowledge transfer from UICs is a transparent and empathetic perspective of all parties' motivation for the research. Understanding these motivations will promote the development of a project strategy that maximizes KPIs for both parties. Considering that the industry partner is typically the primary funder/co-founder of a majority of UIC projects, reducing the commercialization speed of academic research is more beneficial for the industry and public sector partner. This perspective commonly puts undue pressure on the academic partner, thus deteriorating the relationship. However, should this perspective of short-term gains be leading the tempo of the research? What other methods/perspectives could drive/accelerate the collaboration? An essential theme behind most public policies that promote UICs is the ability to conduct research and commercialize products that neither party could perform independently [8]. That said, the short-term expectations of industry partners may hinder a UIC's economic potential for near-sight gains. This temporal component plays an even more significant role in cybersecurity research due to the rapid advancement and acceleration of the industry itself. Seeing both sides of this temporal coin is critical for the success and sustainability of UICs that result in effective commercialized products. Based on these points, it is apparent that many confounding factors affect go-to-market speed, such as technical competence, company/university features and the relationship therein [14]. The ability to generate a one-size fits all recommendation is truly dependent on all parties: the public sector, the academic institution and the industry partner. It should also be clear that *compromise* significantly affects commercialization speed [15], [16]. Although it is difficult to ascertain the right approach, tailored strategies need to be developed based on specific characteristics of the parties within UICs, focusing on maximizing KPIs. In this spirit, the following strategies can be used to help accelerate knowledge transfer for cyber-security UICs:

*Alignment of research objectives with industry partner strategic objectives [11]*. The industry stakeholders must have a sound understanding of the scientific method and rigor required to generate academic publications. In contrast, academic partners need to be cognizant of the deliverables and requirements of the industry partner.

*Engage highly qualified personnel (HQP) outside the core technical competencies required by the UIC*. This is especially important with cyber-security research, where the substantial technical ceiling requires projects to be led by individuals with strong technical competencies and less business acumen. Engagement with these other parties will help expand the transparency and outreach of the project while making it more accessible to a wider audience.

*Develop a project strategy that parallelizes the research and industry objectives.* For many UICs, work happens sequentially and flows from the academic to the industry partner. This method is difficult to manage and scale, considering the agile nature of academic research and industry objectives. By parallelizing this workflow, the academic partner will be more consistent in generating deliverables, and the industry partner will be more prepared to accept them, all while driving increased engagement.

*Align documentation standards between the organizations.* For most security projects, there are set standards and methods through which both organizations operate. For example, most security software development companies follow an agile methodology and have expected technical documentation approaches (i.e., README.md in repos, Confluence pages, etc.). Having the academic output fit more appropriately with these industry templates accelerates knowledge transfer and increases the accessibility of the research to individuals outside of the UIC.

*Research outputs should be accessible and in a readily ingestible form.* A considerable majority of security research can leverage collaborative technologies to iterate and disseminate information (e.g., git, package managers, etc.). By leveraging these tools (i.e., pip modules, shared GitHub repos, etc.), the engagement and utility of these deliverables are increased. However, these tools are typically an afterthought in UICs, resulting in considerable overhead to leverage existing work.

In addition to strategies specific to the academic and industry partners, the public bodies also play a significant role in knowledge transfer. Considering that UICs are typically co-sponsored by governmental agencies or provide some form of financial relief, the policies they enact considerably affect the adoption and rate of knowledge transfer. Thus, the public sector should emphasize collaborations with clear track records of generating deliverables (i.e., patents, products, jobs) and a motivation to increase the transparency of the work.

## 2.2 Different development cycles in academia and industry: slow, peer-reviewed processes versus ad hoc development - how could methodologies be aligned?

As mentioned during the webinar, one of the key challenges in aligning research in cybersecurity between academia and industry is the disconnect in the development cycles of the two ecosystems. In academia, the research methodology is focused on slow, peer-reviewed processes with a long-term time horizon for commercialization. On the other hand, the industry is focused on more immediate market-driven cybersecurity challenges. Because of the inherent difference in these methodologies, the commercialization of research in academia and industry does not align, and many potential cyber solutions are not brought to the market by the industry. It is, therefore, essential to ensure that the correct research problems are defined at the start of any collaborative exercise between industry and academia. Such a collaborative exercise should also include feedback from all stakeholders, including potential customers. There should also be a process of carrying out a Return on Investment (RoI) analysis, as this is critical to the success of the commercialization of the technology.

A great case study for better collaboration between industry and academia is the research carried out in the new and emerging field of automotive cybersecurity. "Automotive Cybersecurity" is the application of cybersecurity principles and best practices to the ecosystem of "Connected & Autonomous Vehicles (CAVs) / Electric Vehicles (EVs)". This transformation in the automotive environment has generated new and unknown cybersecurity challenges that are not understood by the wider cybersecurity community, whether in academia or industry. This is because Automotive Cybersecurity is different from IT Cybersecurity [17] and needs to be addressed for research purposes. However, that entails a paradigm shift in the perception of cybersecurity applicable in an IT environment to the application of cybersecurity for automotive technologies such as CANbus and ECUs in vehicles.

Currently, the development of cybersecurity solutions in the automotive industry is being driven by market needs and spearheaded by the industry itself. Standards such as ISO 21434 Road Vehicles Cybersecurity Engineering Standard [18] have been developed by Industry Cybersecurity leaders to foster cybersecurity application in the automotive environment. Furthermore, Government mandates such as the UNECE WP.29 R155 [19] and R156 [20] are also enabling the implementation of cybersecurity in the automotive environment. These market-driven cybersecurity enablers have incentivized the cybersecurity industry to provide solutions which address myriad challenges in automotive cybersecurity. However, academia is still not aligned with the industry in addressing these myriad challenges posed by the transformation in the automotive industry. Based on its research methodology, industry is still focused on traditional IT Cybersecurity challenges, and there is only limited research in this critical and new area of cybersecurity. Academia needs to align with the needs of the industry and work closely to meet these cybersecurity demands as per the automotive Industry. There is a need to enable a Joint Innovation Program for Automotive Cybersecurity. Entities such as the CSA Group have recently brought together a Working Group on CAV Cybersecurity, Privacy and Data Management and published the gaps identified by these industry leaders [21]. Research can be carried out in these areas to further enable tools and technologies in Automotive Cybersecurity. By applying these principles of Cybersecurity Risk, Compliance and Governance in the Automotive environment, academia and industry can come together to add value to this critical sector which is an inherent part of our daily lives. Cyber Governance in the Automotive environment has to address the domains of People, Processes and Technology from a new dimension [22]. Such a timely collaboration between industry and research would ensure that our future roads are safe & secure.

## 2.3 General Knowledge transfer: how to connect the knowledge silos and different approaches to knowledge transfer in industry and academia?

Universities and industry are organizations that both value knowledge. Knowledge discovery and research excellence are among the existential objectives of universities. Modern industrial organizations also use knowledge to achieve their financial objectives. Although collaboration between academia and industry comes naturally, its realization follows with challenges and impediments. In this webinar, experts from the industry and academia discussed the different characteristics and methodologies of knowledge transfer with a focus on cybersecurity from both perspectives. The following recommendations were proposed to overcome the challenges both parties face in university-industry knowledge transfer and mobilization.

*Knowledge Transfer*: Generally, the term knowledge transfer (KT) is used to express any exchange of knowledge between individuals and organizations. Bloedon et al. [23] define KT as a process by which knowledge related to "making or doing of useful things" in one organization is transferred and brought into use within another organization. In the context of university-industry collaboration, Friedman et al. [24] define KT as "the process whereby invention or intellectual property from academic research is licensed or conveyed through use rights to a for-profit entity and eventually commercialized." For this process to be realized, several activities such as teaching, interaction management, sharing data and technology need to be completed. Alignment between these activities, also called "KT practices or methodologies," is necessary for successful knowledge mobilization between organizations. The panelists identified the following contrasting views on cybersecurity knowledge creation and transfer between university and industry.

*Knowledge Discovery vs Business Outcomes*: The panel acknowledged that universities and industries have different primary outcomes that create barriers to knowledge transfer [25, 26]. The primary objectives of universities are knowledge discovery and innovation, and the outcomes of research are disseminated to the research community and public through scholarly publications. While knowledge discovery is also important for industry for product development and innovation, the key focus of industry is to achieve business outcomes. Industry has multiple dimensions and stakeholders, so the commercialization of academic research touches many areas beyond the scope of the academic partner. For example, twenty years ago, privacy researchers had no common intersection with designers of network security solutions who viewed the data being transmitted. As a result, there is a disconnect between academics researching cutting-edge problems in specific areas that are irrelevant to the industry partner. In fact, there is no common language between universities and industry, which makes knowledge transfer challenging.

*Operational Effectiveness*: A key component to accelerate knowledge transfer is to align and optimize the internal drivers (operational effectiveness) of universities and industry [25, 27]. The panel discussed how universities and industry could align their internal drivers based on a common goal to collectively solve problems (defined as the forcing function by the panel) [28]. When academics and industry agree on a well-defined problem based on the needs of the industry, this agreement could lead to the transfer of knowledge, directly assisting the industry partner [29]. Poorly defined problems, on the contrary, inhibits the transfer of knowledge, because the outcomes of research are not practical or cannot be communicated with the target audience. While defining the problems currently faced by the industry is important, academics must also motivate their industrial partner on the long-term potential of their research to give the company a future competitive advantage (e.g., the potential of homomorphic encryption in data mining). Industrial partners need to collaborate with universities in parallel with the product development processes instead of taking over after research is completed. Another proposal by the panel was to incorporate internal drivers for the industry into the collaboration measures of success [23]. For example, research proposals would link specific industry outcomes to the release of funds. The review of research proposals could also include points/ranking based on academics' success in mobilizing research from lab to industry, such as patents, new products, and estimated revenue.

*Common Language Between University and Industry:* It is important to find a common language and/or theme that universities and industry can rely on by overlapping the competitive advantages from both sides [30]. For example, the main motivation for the industry to collaborate with universities is the opportunity to connect with universities' expertise, the courses they are teaching, and the main research topics being pursued [26]. The panel suggested that universities open slots in relevant courses (e.g., advanced cryptography) to improve communication between academics and industry staff. Researchers should also visit the company (lunch and learn) to explain to the staff their research in a non-technical way. The industry must work to expose the academic partner to the various stakeholders and perspectives within the company. Academics would benefit from understanding how their research is industrialized. The technical staff at the company would also be able to explain their requirements to the academic partners (e.g., documentation) to make the transferred knowledge more accessible to the industry. In summary, the panel had the following recommendations:

- Academics and industry must work together to clearly define the problem instead of only aligning primary outcomes.
- Industry collaborators should consider evaluating research proposals based on researchers demonstrating a track record of knowledge transfer from university to industry.
- Universities should leverage their teaching capability to assist industry partners in understanding the academic topics being investigated.
The industry should incorporate academics into their development process so academics understand the requirements for their research to be industrialized.

**2.4 How to successfully communicate your commercialization plans between academia and industry**

To foster communication of commercialization plans between academia and industry, prevalent differences between the two units in terms of their key objectives, work processes, and expected outcomes that may impede their collaboration as a coherent unit should be carefully studied.

*Having a clear statement on problems and potential growth opportunities*: To enable the commercialization of research outcomes and technology innovations, academia should propose a clear definition of a problem statement to provide an understanding of the relevant industry problem. Providing a clear assessment of the target market, including the market size, targeted segments with growth potential and opportunities, as well as the competitive advantage that would be gained in the future, may help the industry adopt the research outcomes and innovation generated by academia. On the industry side, while creating a partnership and innovation programs, posing similar questions to clarify the problems and expected achievements, such as: "what do we try to accomplish as a company?", "Where do we have the gaps?", "Which of these knowledge gaps can be refined as opportunities for growth?" would be helpful. Once the industry has clear, refined, and easy-to-understand answers to these fundamental questions, finding and adopting the novel and innovative solutions offered by academia would be possible. As knowledge transfer from one form to the other is indeed challenging, providing clear answers to these questions at earlier stages would help transform knowledge to the right form [31] and make it available to consume by the other party.

*Establishing channels to enable communication and collaboration between industry and academia:* Both academia and industry should consider ways to create space for knowledge transfer. Hiring coop students on the industry side is one of the examples that a mechanism for interaction and collaboration could take place, which would eventually assist with the transfer of knowledge as well as the development of a shared understanding, knowledge, and culture. However, these efforts should not be limited to the industry's efforts. The academic culture is often driven by a "publish or perish" culture, which doesn't support the motivation to create space for commercialization [32]. Academia should revise their promotion systems to incentivize academics to expand their research context and be motivated to contribute to industry projects or commercialize their findings.

*Recognizing that commercialization is a multistage process:* Commercialization is a multi-stage process that requires the active involvement of multiple stakeholders [33, 34, 35]. The first step includes translating research into function, which can serve as a foundation for bringing a new product or service to a given market. The validation of a function being able to convert that to a new product or service should be followed by the validation of the market of that product or service. So, the researcher should not only validate the function but also whether this function is commercially viable given a specific market. Market validation helps the researcher understand whether the produced function is important and valuable enough to be invested in and adopted by the industry, as not all research may be worth commercializing. A critical assessment of the characteristics of the market is important to understand whether the commercialized product can serve the market's needs and whether the size of the market it is delivered to can make the efforts worthwhile and profitable.

*Recognizing the role of incubators and futurists in this process*: Incubators are one of the intermediary institutions that could support and facilitate entrepreneurship through their administrative and consulting services and strong ties to business networks [36]. However, currently, there is a gap within the cybersecurity innovation ecosystem such that most of the focus on incubators and accelerators lies primarily on active teaching entrepreneurship. However, these parties should have a supporting role in transferring knowledge among stakeholders that would lead to commercialization. Incubators and accelerators can play an important role in converting research into function and market validation exercises, creating an ecosystem where entrepreneurs, academic researchers, and mentors work together. Futurists can also have a critical role in bridging the gap between industry and academia that hampers the translation of knowledge that may lead to commercialization. In most cases, organizations do not actively seek or assess the future viability of academic research and innovation. Futurists can work on understanding the potential of technology and research innovations that can be commercialized to gain a competitive advantage. That way, they can serve as a layer of the process to help the industry differentiate between the products and services that are worthy of investing in or not. Currently, in the cybersecurity ecosystem, the founders aiming to innovate radically usually encounter resistance from investment communities because of their interest in short or midterm benefits. However, a deep tip ecosystem like cybersecurity needs futuristic technologies, such as algorithmic technologies, yet currently, it's hard to attract any commercial interest to invest in these technologies. Hence, futurists should advocate for the products and services that are radical in their innovation to benefit the industry in the long term.

*Acknowledging the core differences between academia and industry:* Academia and industry have fundamental differences in key aspects, including their core objectives, motivations, expected outcomes, promotions systems, etc. Since the core motive of academia is not producing a commercial product, or commercialization of generated knowledge, technology, or innovation, bringing academia and industry together in the same "commercialization" pot would be a futile effort. Instead, both parties should target innovation by solving the right problems, bringing the right people, and developing a shared language. Another solution could be supporting the existence of other intermediary institutions, such as business innovation centers within universities, to create a conducive context that collocates similar activities conducted by academic and industry entrepreneurs to create synergies and spillover effects [37, 38], and thus enable the translation of novel technology from lab to market [39]. Commercialization would follow as the product of this mutual effort.

## 3. Conclusions

In this paper, we have provided an overview of the most critical gaps and solutions identified by an interdisciplinary expert exchange on bridging the bubbles between academia and industry in cyber security research. We focused on four primary questions 1) the speed of transition from industry to/from academics as related to R&D solutions, 2) the alignment of the development cycles, peer-reviewed processes versus ad hoc development, 3) knowledge transfer, mobilizations, and dissemination format of research, and 4) strategies for communicating commercialization plans between both parties. Experts from the industry and academia discussed different insights, methodologies, and strategies that can connect the gap in academia and industry with a focus on cybersecurity from both perspectives.

## 4. Declarations:

- Availability of data and material: The link to the webinar is available at:

https://cybersecurecatalyst.ca/watch-bridging-the-bubbles-connecting-academia-and-industry-in-cybersecurity-research/
- **Competing interests:** N/A
- **Funding:** Rogers Cybersecure Catalyst
- **Authors' contributions:** All Authors contributed Equally.

- **Acknowledgements:** The authors acknowledge funding from the Rogers Cybersecure Catalyst at Toronto Metropolitan University. We would like to thank the external panelists, Shiladitya Sircar and Sumit Bhatia, the catalyst fellows, directors and administration for their contribution and participation to the webinar.